\documentclass{aa}

\usepackage{graphicx}
\usepackage{wasysym}

\def\comment#1{\noindent{\bf }}

\def\wig#1{\mathrel{\hbox{\hbox to 0pt{%
          \lower.6ex\hbox{$\sim$}\hss}\raise.4ex\hbox{$#1$}}}}

\def\mjup{$\rm M_J$}
\def\rjup{{\small$\rm\overline{R}_J$}}

\def\teff{T_{\rm eff}}
\def\teq{T_{\rm eq}}
\def\tabs{T_{\tiny\RIGHTcircle}}
\def\labs{L_{\tiny\RIGHTcircle}}

\def\lint{L_{\rm int}}
\def\tint{T_{\rm int}}
\def\tday{T_{\rm day}}
\def\tnight{T_{\rm night}}

\def\m{\rm\, m}

\def\ergscm{\rm\,erg\,s^{-1}\,cm^{-2}}
\def\W{\rm\,W}

\def\km{\rm\,km}
\def\cm{\rm\,cm}

\def\erg{\rm\, erg}
\def\hotjups{Pegasi planets}    % plural noun  (hot Jupiters)
\def\s{$\,\,$}                 % space

\def\apj{ApJ}
\def\apjs{ApJS}
\def\apjl{ApJL}
\def\nature{Nature}
\def\aa{A\&A}
\def\icarus{Icarus}

\def\jgr{J. Geophys. Res.}
\def\science{Science}

\begin{document}

%\baselineskip=24truept

\thesaurus{08}
\title{Evolution of ``51\,Peg\,b-like'' Planets}
\author{Tristan Guillot\inst{1}
\and Adam P.\,Showman\inst{2}}
%\offprints{T. Guillot}
\institute{Observatoire de la C\^ote d'Azur, Laboratoire Cassini, CNRS
UMR 6529, 06304 Nice Cedex 4, France; guillot@obs-nice.fr
\and University of Arizona, Department of Planetary Sciences and Lunar and Planetary
Lab, Tucson, AZ 85721; showman@lpl.arizona.edu}
\date{Submitted 23 March 2001; Accepted 7 November 2001}
%\date{Submitted March 23, 2001; Revised November 2, 2001}
\titlerunning{Evolution of \hotjups}
\authorrunning{T. Guillot and A.P. Showman}
\maketitle

\begin{abstract}
About one-quarter of the extrasolar giant planets discovered
so far have orbital distances smaller than 0.1\,AU.
These ``51\,Peg\,b-like'' planets can now be directly characterized,
as shown by the planet transiting in front the star HD209458. 
We review the processes that affect their evolution.

We apply our work to the case of HD209458b, whose radius has been
recently measured. We argue that its radius can be reproduced only
when the deep atmosphere is assumed to be unrealistically hot.  When using
more realistic atmospheric temperatures, an energy source appears to
be missing in order to explain HD209458b's large size.
The most likely source of energy available is not in the
planet's spin or orbit, but in the intense radiation received from the
parent star.  We show that the radius of HD209458b can be reproduced
if a small fraction ($\sim1\%$) of the stellar flux is transformed
into kinetic
energy in the planetary atmosphere and subsequently converted to
thermal energy by dynamical processes at pressures of tens of bars.
\end{abstract}

\section{Introduction}

The detection of planetary-mass companions in small orbits around solar-type stars
has been a major discovery of the past decade.  To date, 73 extrasolar
giant planets (with masses $M\sin i<13\,$\mjup, \mjup\ being the mass
of Jupiter and $i$ the inclination of the system) have been detected by
radial velocimetry.  
Fifteen of these (21\%) have distances less than 0.1 AU,
and ten (14\%) have distances less than 0.06 AU (see Marcy et al.
\cite{Mar00} and the discoverers' web pages).  This is for example
the case with the first extrasolar giant planet to have been discovered,
51 Peg b (Mayor \& Queloz \cite{MQ95}). These close-in planets
form a statistically distinct population: all planets with
semi-major axis smaller than 0.06 AU have near-circular orbits while
the mean eccentricity of the global population is $<e>\approx 0.27$. This is
explained by the circularization by tides raised on the star by
the planet (Marcy et al. \cite{Mar97}). One exception to this rule,
HD83443b ($e=0.079\pm 0.033$),
can be attributed to the presence of another eccentric planet
in the system (Mayor et al. \cite{May01}). As we shall see, the
planets inside $\sim$0.1 AU also have very specific properties due to the
closeness to their star and the intense radiation they receive.
For this reason, following
astronomical conventions, we choose to name them
after the first object of this class to have been discovered:
``51\,Peg\,b-like'' planets, or in short ``Pegasi planets''.

Such planets provide an
unprecendented opportunity to study how intense
stellar irradiation affects the evolution and atmospheric circulation
of a giant planet.  Roughly 1\% of stars surveyed 
so far bear \hotjups\s in orbit, suggesting that they are not
a rare phenomenon.
Their proximity to their stars increases the likelihood
that they will transit their stars as viewed from Earth, allowing a
precise determination of their radii. (The probability 
varies inversely with the planet's orbital
radius, reaching $\sim$10\% for a planet at 0.05 AU around a solar-type
star.)  One planet, HD209458b, has already been observed to 
transit its star every 3.524 days (Charbonneau et al. \cite{Cal00}; 
Henry et al. \cite{Hal00}).  The object's mass is $0.69\pm0.05$\,\mjup\, ,
where \mjup$=1.89\times 10^{27}$\,kg is the mass of
Jupiter.  Hubble Space Telescope measurements of the transit
(Brown et al. \cite{Betal01})
imply that the planet's radius is $96300\pm4000$\,km.
An analysis of the lightcurve combined with atmospheric models shows
that this should correspond to a radius of 94430\,km at the 1\,bar level
(Hubbard et al. \cite{Hub01}).  This last estimate corresponds to
1.349\,\rjup, where \rjup$\,\equiv 70,000\km$ is a characteristic 
radius of Jupiter.
This large radius, in fair agreement with theoretical predictions
(Guillot et al. \cite{Gui96}), shows unambiguously that HD209458b
is a gas giant.

We expect that the evolution of \hotjups\s depends more on the
stellar irradiation than is the case with Jupiter.
HD209458b and other \hotjups\s differ qualitatively from Jupiter
because the globally-averaged stellar flux they absorb is 
$\sim10^8\erg\cm^{-2}$ ($10^5\W\m^{-2}$), which is $\sim10^4$ times
greater than the 
predicted intrinsic flux of $\sim10^4\erg\cm^{-2}$.  (In contrast,
Jupiter's absorbed and intrinsic fluxes are the same within a factor
of two.)    Several evolution calculations
of \hotjups\s have been published (Burrows et al. \cite{Bur00},
Bodenheimer et al. \cite{Bod01}), but these papers disagree about
whether HD209458b's 
radius can be explained, and so far there has been no general discussion of
how the irradiation affects the evolution. Our aim is to help fill this gap.  

Here, we quantify how atmospheric processes affect the evolution 
of \hotjups\s such as HD209458b.  First (Section 2), 
we show that the evolution is sensitive to the assumed atmospheric 
temperatures.   This sensitivity has not previously been documented,
and quantifying it is important because the temperature profiles
appropriate for 
specific planets remain uncertain (e.g., no atmospheric radiative transfer 
calculation for HD209458b yet exists).  Our works suggests that
the discrepancy between the  predictions of Burrows et
al. (\cite{Bur00}) and  Bodenheimer et al. (\cite{Bod01}) can be
largely explained by their different assumptions about atmospheric
temperature.

Second, the effect of atmospheric dynamics on the evolution has to date been 
neglected.  For example, current models assume the day-night temperature  
difference is zero, despite the fact that substantial day-night temperature 
variations are likely.  In Section 3.1 we demonstrate how the evolution
is modified when a day-night temperature difference is included.
Furthermore, the intense stellar irradiation will lead to production of
atmospheric kinetic energy, and transport of this energy into the interior
could provide a substantial energy flux that would counteract the loss
of energy that causes planetary contraction.  In Section 3.2 we 
investigate this effect.

The research has major implications for HD209458b.
Early calculations implied that \hotjups\s contract
slowly enough to explain HD209458b's large radius (Guillot et
al. \cite{Gui96}, 
Burrows et al. \cite{Bur00}).  But recent calculations of irradiated
atmospheres suggest that the actual deep atmosphere is colder than 
assumed (Goukenleuque et al. 2000).  When such realistic temperatures 
are adopted (our Section 2), the planet 
contracts too fast and the
radius is $\sim0.2$--0.3\,\rjup\ too small.  Bodenheimer et al. (\cite{Bod01})
argued that tidal heating from circularization of the orbit would slow
the contraction, leading to a larger radius, but this is a transient
process that would end $\sim 10^8$ years after the planet's formation.   
Instead we argue that kinetic energy 
produced in the atmosphere is transported into the interior and dissipated 
(Section 3.2). We show that plausible downward energy fluxes can 
slow or even halt the planet's contraction, 
allowing HD209458b's radius to be explained.

In a joint paper (Showman \& Guillot \cite{SG01}, Paper II) we consider
the atmospheric dynamics 
of these planets, with emphasis on how the atmospheres respond
to stellar heating and gravitational tidal interactions, and
on the observable consequences.

\section{Sensitivity of Evolution to Atmospheric Temperature}

The upper boundary condition of evolution models consists of
a relationship between the effective
temperature and some deeper temperature (say that at 10 bars) to which
the model's interior temperature profile is attached.
Here we show that the evolution is sensitive to the assumed relationship
(i.e., to the assumed atmospheric temperature structure).  

Before we begin, we provide some definitions.
We define the effective temperature of the irradiated planet as
\begin{equation}
4\pi R^2\sigma\teff^4 = \labs+\lint,
\end{equation}
where $R$ is the planet's radius, $T_{\rm eff}$ is its effective
temperature, $\sigma$ is the Stefan-Boltzmann constant,
$\labs$ is the part of the stellar luminosity absorbed by the
planet and $\lint$ is the intrinsic luminosity of the planet due to
its cooling and contraction (and possibly other processes such as
radioactivity or thermonuclear reactions in the case of massive
objects).

The temperature corresponding to the intrinsic
planetary flux, called the ``intrinsic'' temperature $\tint$, is defined by
\begin{equation}
4\pi R^2\sigma\tint^4 = \lint.
\end{equation}
Similarly, we define $\tabs$ from the absorbed stellar luminosity
$\labs$.
$\tabs$ is the effective temperature towards which the planet tends as
it cools and $\lint$ diminishes (see e.g. Hubbard
\cite{Hub77}). It is a function of the
Bond albedo (i.e. the ratio of the luminosity directly reflected to
the total luminosity intercepted by the planet).
$\tabs$ can be viewed as the effective temperature reached by the
planet after it has lost its internal heat, and is hence sometimes
noted $\teq$ (e.g. Guillot et al. \cite{Gui96}; Saumon et
al. \cite{Sau96}).

\subsection{Atmospheric boundary conditions}

\begin{table*}[htb]
\caption{Parameters used in evolution models}
\label{tab:param}
\begin{tabular}{lll}
\hline\hline
Parameter & Value & References/remarks \\ \hline
Evolution model & CEPAM & Guillot \& Morel (\cite{GM95}) \\
\hline
Mass & $M=0.69\rm\,M_{Jup}$ & ($\rm M_{Jup}\equiv 1.89\times 10^{30}\,g$) \\
Absorbed stellar heat & $\tabs=1400\,$K & \\
Radius & $R=1.35\rm$\,\rjup & (\rjup$\equiv 7\times
10^{9}\,$cm)\\ \hline
EOS & ``interpolated'' & Saumon, Chabrier \& Van Horn (\cite{Sau95}) \\
Helium mass mixing ratio & $Y=0.30$ & \parbox{8cm}{Higher than solar
in order to mimic a solar abundance of heavy elements} \\
Rosseland opacities & --- & Alexander \& Ferguson (\cite{AF94})
incl. interstellar grains \\
Rotation & 0 & Neglected \\
Core mass & 0 & Not considered \\
Atmospheric boundary & --- & \parbox{8cm}{From Marley et
al. (\cite{Mar96}); Burrows et al. (\cite{Bur97}) \\See
Eqs.~(\ref{eq:T-std}) and (\ref{eq:T-cold})} \\
\hline
\end{tabular}
\end{table*}

We consider two evolution models of
HD209458b based on the parameters listed in table~\ref{tab:param};
the two models differ only in their prescription for the
atmospheric boundary condition.   

\begin{figure}[ht]
\begin{center}
\resizebox{\hsize}{!}{\includegraphics[angle=0]{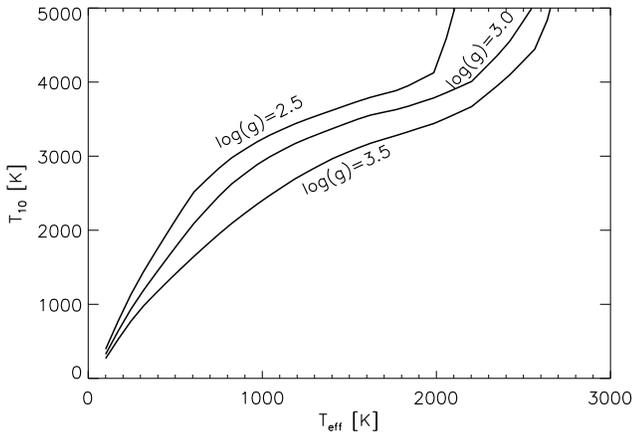}}
\caption{%
Surface boundary condition (temperature at the 10 bar level) that
has been used in several published evolution models, and which we
dub the ``hot'' case, as a
function of effective temperature for three different gravities:
$log(g)$=2.5, 3 and 3.5. ($g$ is in units of $\rm cm\,s^{-2}$).
(See Saumon et al. (\cite{Sau96}) for a discussion.)
}
\label{fig:te_t10}
\end{center}
\end{figure}

Our first evolution sequence, dubbed the ``hot'' case, uses the 
standard boundary condition from Guillot et al. (1996) and Burrows et
al. (2001a).   These papers adopted an atmospheric 
structure of an {\it isolated} object with the expected
effective temperature, which provides a fair
fit to the evolution of Jupiter.  The surface boundary
condition consists of a relationship between the temperature at the 10\,bar
pressure level $T_{\rm isolated}$, the effective temperature $\teff$ and the
gravity $g$ of an isolated planet/brown dwarf derived by several
authors (see Marley et al. \cite{Mar96}; Burrows et
al. \cite{Bur97}): 
\begin{equation}
T(P=10\ {\rm bar})=T_{\rm isolated}(\teff,g)
\label{eq:T-std}
\end{equation}
This approximation is exact in the limit when the stellar flux
is {\it entirely} absorbed at the 10\,bar
level, or in a deep adiabatic (convective) region, as is the case for
Jupiter (see Hubbard \cite{Hub77} for a detailed discussion of the
effect of insolation on Jupiter's evolution). 
Figure~\ref{fig:te_t10} shows the variation of $T_{\rm isolated}$ with 
effective temperature and gravity, for values of interest in the case
of \hotjups.  

Unfortunately, the approximation becomes incorrect in the case of
strongly irradiated planets because of the growth of a thick external
radiative zone. Another boundary condition has
therefore to be sought: either part of the stellar flux is able to
penetrate to deeper levels ($P_0>10$\,bar) and lead to a boundary
condition defined by $T(P_0)>T_{\rm isolated}$, or most of the stellar
flux is absorbed at $P_0<10$\,bar, yielding $T(P_0)<T_{\rm
isolated}$. (This is due to the fact that in the radiative zone
$dT/dP\propto F$, where $F$ is the flux to be transported).
It will be shown hereafter (see section~\ref{sec:flux}) that
Eq.~\ref{eq:T-std} is effectively an upper limit to the boundary
temperature because continuum opacity sources only effectively limit
the penetration of the stellar photons. 
Indeed, more detailed models of the atmospheres of \hotjups\s
have shown that most of the starlight is absorbed at pressures less
than 10 bar, and that Eq.~\ref{eq:T-std}
overestimates the atmospheric temperatures by as much as 
300 to 1000\,K (Seager \& Sasselov \cite{SS98}, \cite{SS00};
Goukenleuque et al. \cite{Gouk00}; Barman et al. \cite{Bar01}). 
%This is mainly because most of the starlight is absorbed at
%pressures less than 10 bars, in a region which is generally
%radiative: the flux transported
%to deeper levels is then smaller, and so is the radiative temperature
%gradient (or lapse rate, in absolute sense). As a consequence, shallow
%absorption of the stellar flux leads to higher temperatures at low
%pressures, but lower temperatures at deeper levels than compared to
%the deep absorption case. 

Because these atmospheric models do not presently span the effective
temperature and gravity range that is needed, and more importantly
because they assume unrealistic intrinsic temperatures, we chose
to construct an arbitrary boundary condition based on the results of
the isolated case. For a given $\teff$, the isolated case provides an
upper bound to the ``surface'' temperature and by extension to the
temperatures in the planetary interior. In order to have an approximate lower
bound that agrees with atmospheric models of irradiated giant planets,
we assume (i) a lower value of $P_0=3$\,bar, and (ii) that the 
temperature at that level is given by:
\begin{equation}
T(P=3\ {\rm bar})=T_{\rm isolated}(\teff,g)-1000\,\rm K.
\label{eq:T-cold}
\end{equation}
Evolution calculations made with this boundary condition are dubbed
the ``cold'' case.

Note that we found {\it a posteriori} that the choice of $P_0$ is
almost unconsequential for the evolution calculations. This is because
the external radiative region quickly becomes almost isothermal (see
fig.~\ref{fig:prof_hotcold} hereafter). However, the consequences of the
cooler temperatures are profound, and as we shall see lead to a much
faster evolution.

In this context, Bodenheimer et al. (\cite{Bod00};
\cite{Bod01}) assume that the temperature at optical depth $2/3$
(corresponding in their model to a pressure of the order $\sim
1$\,mbar) is
equal to the effective temperature $\teff$, an approximation that
leads to an 
underestimation of the actual atmospheric temperatures. As a
consequence, their 1 bar temperatures are of the order of $\sim
1400$\,K, 
i.e. even lower than what is implied by Eq.~(\ref{eq:T-cold}). 
This would imply an extremely inefficient penetration
of the stellar flux in the planetary atmosphere, in disagreement with
detailed models of these atmospheres. 
We therefore prefer to use Eq.~(\ref{eq:T-cold}) as our ``cold''
boundary condition.

\subsection{Evolution models of HD209458b}

The evolution of HD209458b is calculated as described in Guillot et
al. (\cite{Gui96}), using the parameters given in
Table~\ref{tab:param}.
Because of the high stellar insolation, the contraction and cooling of
the planet from a high entropy initial state is only possible through
the build-up and growth of a radiative zone (Guillot et al. \cite{Gui96}).

\begin{figure}[ht]
\begin{center}
\resizebox{\hsize}{!}{\includegraphics[angle=0]{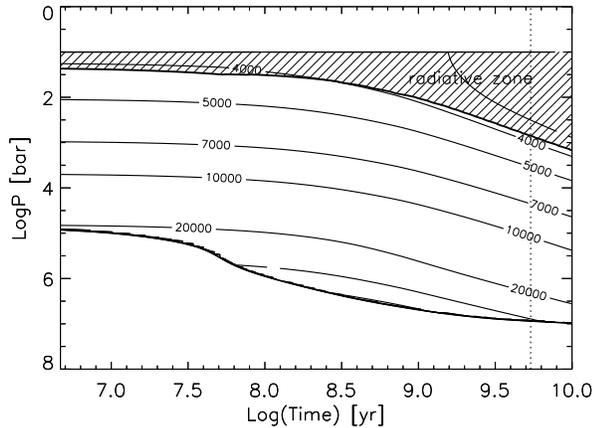}}
\caption{%
Evolution of HD209458b using the ``hot'' atmospheric boundary
condition (Eq.~(\protect\ref{eq:T-std})).  The evolution of the central
pressure with time
is shown as the bottom thick line. The planet is convective except for
an upper radiative zone indicated by a hashed area. Isotherms from
4000 to 20\,000\,K are indicated. The isotherms not labelled correspond
to 3500, 30\,000 and 40\,000\,K. The dashed line indicates the time
necessary to contract the planet to a radius of 1.35\,\rjup. [$\rm
1\,bar=10^6\,dyn\,cm^{-2}$].
}
\label{fig:ev-std}
\end{center}
\end{figure}

\begin{figure}[ht]
\begin{center}
\resizebox{\hsize}{!}{\includegraphics[angle=0]{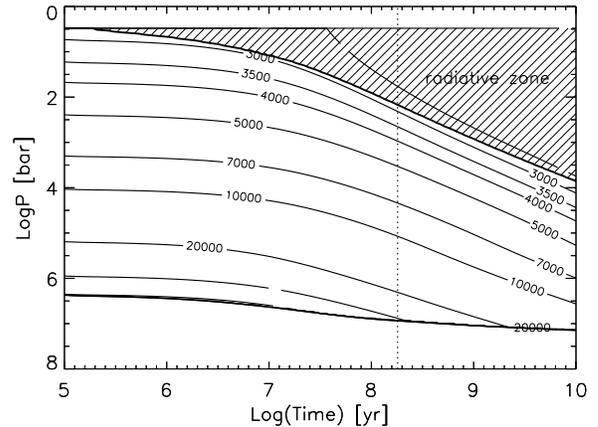}}
\caption{%
Same as Fig.~\protect\ref{fig:ev-std} but with the ``cold'' boundary
condition given by Eq.~(\protect\ref{eq:T-cold}). Unlabeled isotherms
are for $T=2500$, 30\,000 and 40\,000\,K.
}
\label{fig:ev-cold}
\end{center}
\end{figure}

The evolution of the interior of HD209458b for the two cases is shown in
figures~\ref{fig:ev-std} and ~\ref{fig:ev-cold}. After a rapid
contraction during which both the central pressure and temperature
increase, the onset of degeneracy leads to a cooling of the interior
as the planet continues to contract. The cooling of the interior
proceeds despite the fact that the atmospheric temperatures remain
nearly constant thanks to the growth of a radiative region in the
planet's upper 
layers, as indicated by the dashed area.

In the case of the ``hot'' atmospheric boundary condition
(Fig.~\ref{fig:ev-std}), the measured radius (1.35\,\rjup)
is attained after 5.37\,Ga ($5.37\times 10^9$ years), which is
compatible with the age of the G0 star HD209458
(see Burrows et al. \cite{Bur00}). The radiative zone then extends to
about 730\,bar, and the intrinsic luminosity is $7.7\times
10^{24}\rm\,erg\,s^{-1}$  (2.3 times that of Jupiter), which
corresponds to an intrinsic temperature of 105K.

The ``cold'' atmospheric boundary condition (Fig.~\ref{fig:ev-cold})
yields a much faster evolution: the planet then shrinks
to 1.35\,\rjup\ in only 0.18\,Ga (see also fig.~\ref{fig:r_vs_t}
hereafter). This is incompatible with the age 
derived for HD209458b.   The radiative/convective boundary in this
model (at 0.18\, Ga) is at 160\,bar, due to the
higher intrinsic luminosity equal to $1.9\times 10^{26}\rm\,erg\,s^{-1}$,
equivalent to an intrinsic temperature of 234\,K.

\begin{figure}[htb]
\begin{center}
\resizebox{\hsize}{!}{\includegraphics[angle=0]{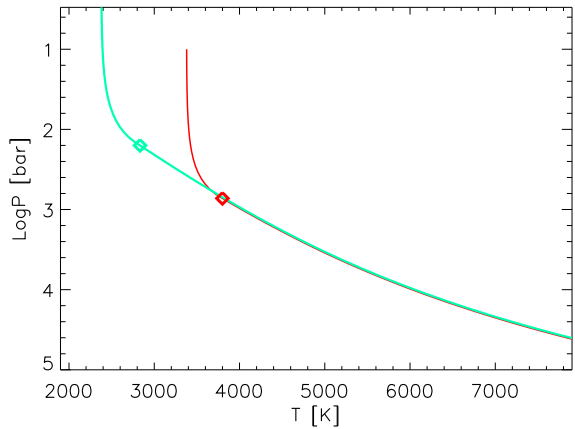}}
\caption{%
Temperature profiles for the 5.37\,Ga-old ``hot'' model
(thin black line) and the 0.18\,Ga-old ``cold'' model (thick grey line),
i.e. when they match HD209458b's measured radius. The diamonds 
indicate the radiative/convective boundary. }
\label{fig:prof_hotcold}
\end{center}
\end{figure}

The fact that the evolution is faster in the ``cold'' case may seem
counterintuitive.  It occurs because the intrinsic luminosity is
proportional to the temperature {\it gradient}, not to the
temperature itself. As shown by Fig.~\ref{fig:prof_hotcold} hereafter,
the temperature variation in the radiative region is more pronounced in
the ``cold'' case than in the ``hot'' case. Basically, this is because
the temperatures at deep levels are fixed by the condition on the
radius, but that the surface temperatures are very different in the
``cold'' and ``hot'' cases. 

Bodenheimer et al. (\cite{Bod01}) obtain radii that slightly exceed
those of our ``cold'' case, despite their lower atmospheric
temperatures.
However, this is probably due to their  
lower assumed value for the helium abundance $Y=0.24$, whereas we chose
$Y=0.30$, a value representative of conditions in the solar nebula and
that accounts for a solar proportion of heavy elements. In both cases,
young ages are required to reproduce the measured planetary
radius.  

We therefore feel that because the ``cold'' atmospheric boundary
condition is preferable to the ``hot'' boundary condition, there is a
problem in explaining HD209458b's radius. An absolute proof of this
statement would require calculations of many different models using
different assumptions, which we will not attempt in this paper. The
conclusion should be relatively secure however, because several
factors point towards a reduction of the planet's radius compared to
what we have calculated: (i) The atmospheric temperatures could be even
lower than in the cold case; (ii) The opacities used include the
presence of abundant grains in the atmosphere and does not account for
their gravitational settling;
(iii) Our choice
of the equation of state tends to yield larger radii than would be the
case using an equation of state that consistently models the
molecular/metallic transition (see Saumon et al. \cite{Sau95}); (iv)
The presence of a central core will tend to greatly reduce the
planet's radius (Bodenheimer et al. \cite{Bod01}). 

An additional source of energy then appears to be required. We note that
the presence of a 
hydrogen/helium phase separation, like in Jupiter and 
Saturn (e.g. Stevenson \& Salpeter \cite{SS77}; Guillot \cite{Gui99}),
is not a valid alternative because of the high interior temperatures
involved in the case of \hotjups\s (e.g. 20,000\,K at 1\,Mbar).

An important aspect of the hot and cold models is shown in
Fig.~\ref{fig:prof_hotcold}: Apart from the
outer radiative layers, the two models possess a very similar interior
structure at the times (5.37 and 0.18 Ga for the hot and cold cases,
respectively) when they match HD209458b's observed
radius. This is easily understood by the fact that the
radiative layer encompasses only a small fraction of the
radius. Most of the contribution to the planetary radius is due
to the convective interior. Fixing the radius is,
for a given equation of state and composition, almost equivalent to fixing
the temperature-pressure profile in the deep interior. HD209458b can
be thought of as a relatively well-constrained convective
core underlying a radiative envelope of uncertain mass and
temperature.

\section{Evolution of \hotjups: ``non-standard'' models}
\label{sec:evolution}

\subsection{Implications of atmospheric day/night temperature variations}

We show in Paper~II that the atmosphere should be significantly hotter on the
dayside than the night side. Here we examine the consequences for
the evolution models.  To do so, we use a toy model that, while simple,
elucidates the important physics.  A full two-dimensional model that would allow 
us to calculate the evolution of \hotjups\s including latitudinal or
longitudinal temperature variations will be left for future work.

Let us assume that the planet can be divided in two hemispheres (night
and day) with two different effective temperatures such that
$\tnight\le\tday$.  When the absorbed stellar energy is fully 
redistributed by advection,
$\tday=\tnight=\tabs$. In all cases, energy conservation implies that
$2\tabs^4=\tnight^4+\tday^4$. Therefore, $\tnight=0$ implies
$\tday=2^{1/4}\tabs$; $\tabs=1400$\,K and $\tday=1500$\,K yields
$\tnight=1272$\,K.

\begin{figure}[ht]
\begin{center}
\resizebox{\hsize}{!}{\includegraphics{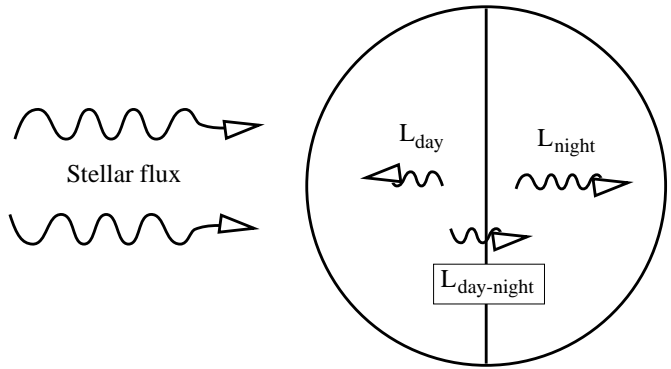}}
\caption{%
Schematics of the day-night toy evolution model. The slow mixing of
the interior leads to a non-radial heat flux $L_{\rm day-night}$ from
the day side to the night side. As a consequence, the intrinsic
luminosity on the day side is smaller and that on the night side
becomes larger.}
\label{fig:twosides}
\end{center}
\end{figure}

Let us do the following gedanken experiment, as illustrated by
Fig.~\ref{fig:twosides}: we suppose that the two
hemispheres cannot exchange energy, and let them evolve from the same
initial state. After a given time
interval $\Delta t$, the central entropy on the day side will have
decreased by a smaller amount than on the night side. This is due to
the fact that a higher atmospheric temperature is equivalent to a
higher stellar flux, and leads to a slower evolution (see Hubbard
\cite{Hub77}; Guillot et al. \cite{Gui95}). In consequence, the night
side will have
become internally colder, have a smaller radius and a larger central
pressure than the day side.

The pressure differences caused by the differential cooling
ensures an efficient mixing
between the two hemispheres on a time scale of decades or less, i.e.
much shorter than the evolution time scale.

We therefore include the effect of atmospheric temperature variations
on the evolution in the following way: We calculate two evolution
tracks of a planet with uniform temperatures $\tnight$ and $\tday$,
respectively. Using these evolution tracks and starting from an
initial condition for which the two models have the same central
entropy, we calculate the entropies of the two sides after a time
interval $\delta t$. We then decrease the entropy of the day side and
increase that of the night side so that both are equal to $(S_{\rm
night}+S_{\rm day})/2$. The process is repeated for each time
step. The cooling of 
the night side is therefore slowed by the mixing of material of
slightly higher entropy from the day side, while the opposite is true
for the day side.

\begin{figure}[ht]
\begin{center}
\resizebox{\hsize}{!}{\includegraphics{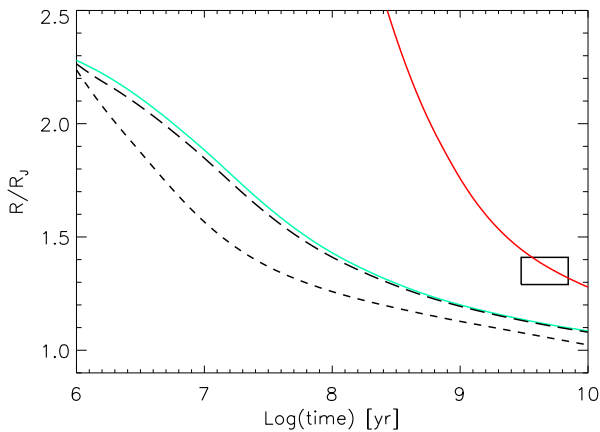}}
\caption{%
Radius of HD209458b (in units of the radius of Jupiter) versus time under
different assumptions. The plain lines correspond to the ``hot'' and
``cold'' evolution cases
shown in figures~\ref{fig:ev-std} (upper curve) and ~\ref{fig:ev-cold}
(lower curve), respectively.
The long dashed line is obtained in the ``cold'' case, when assuming that the
radiative equilibrium effective temperature is 1500\,K on the day side
and 1272\,K on the night side.
The short dashed line is obtained when these effective temperatures are
1664\,K and 0\,K, respectively ($\sim$ no advection). The box
indicates inferred radii and ages of HD209458b.}
\label{fig:r_vs_t}
\end{center}
\end{figure}

The resulting evolution tracks are shown in Fig.~\ref{fig:r_vs_t}.
Not suprisingly, the cooling of an irradiated planet with
inefficient heat redistribution in the atmosphere is faster than if
the stellar heat is efficiently advected to the night side. This is
mainly due to the fact that, with increasing $\Delta T_{\rm
day-night}$, $\tnight$ decreases much more rapidly than
$\tday$ increases, yielding a much faster cooling of the night side.
However, for the temperature variation of 228\,K shown in
Fig.~\ref{fig:r_vs_t} (long dashed line), the effect is limited to a 
variation of $\sim
0.5$\% of the radius after 1\,Ga of evolution or more. The effect is
of course more pronounced if no thermal energy advection occurs in the
atmosphere ($\tnight=0$). In that case (short dashed line), the minimal 
radius is, for a given mass,
composition and stellar insolation, up to 5\% smaller than calculated
in the uniform case.

\begin{figure}[ht]
\begin{center}
\resizebox{\hsize}{!}{\includegraphics{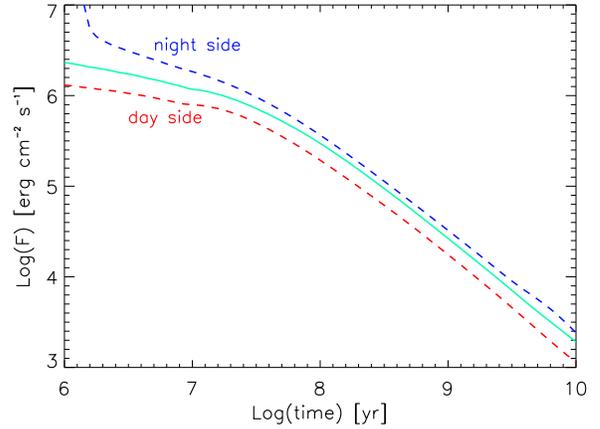}}
\caption{%
Intrinsic planetary fluxes obtained as a function of time in the
``cold'' case with a well mixed atmosphere (plain line). When assuming
that stellar irradiation is imperfectly redistributed over the
planet's atmosphere, the flux on the night side ($\tnight=$1272\,K)
becomes larger than that on the day side ($\tday=$1500\,K).
}
\label{fig:f_vs_t}
\end{center}
\end{figure}

Figure~\ref{fig:f_vs_t} shows that a non-uniform atmosphere has a substantial
effect on the planet's intrinsic flux ($F=L/4\pi R^2$).
After 4.5\,Ga, the flux of our planet with
$\tnight=$1272\,K and $\tday=$1500\,K is 6310 and
2950$\rm\,erg\,cm^{-2}\,s^{-1}$ on the night and day sides,
respectively, which can be compared with the measured intrinsic fluxes
of Jupiter, $5440\rm\,erg\,cm^{-2}\,s^{-1}$ and Saturn,
$2010\rm\,erg\,cm^{-2}\,s^{-1}$ (Pearl \& Conrath~\cite{Pea91}).
This process is analogous to that proposed by Ingersoll \&
Porco (\cite{IP78}) to explain the uniform temperatures of Jupiter.
Stellar insolation tends to suppress the planet's intrinsic heat
flux, and so the planetary heat preferentially escapes in regions
where the insolation is minimal.

\subsection{Evolution with internal energy dissipation}

Current models predict that several-Ga-old \hotjups\s have intrinsic heat
fluxes of $\sim10^4\ergscm$, which is about $10^4$ times less than the
total luminosity of $\sim10^8\ergscm$ resulting from thermal
balance with the stellar insolation.  A fraction $\eta$ of the total
luminosity will be converted into kinetic energy by the atmospheric
pressure gradients.  On Earth, the globally-averaged flux transported
by the atmosphere is about $2.4\times 10^5\ergscm$ ($240\W\m^{-2}$),
while about $2000\ergscm$
is converted into large-scale atmospheric kinetic energy (Peixoto and
Oort \cite{PO92}, p.~385), leading to a value $\eta = 0.01$.  This
energy production can be viewed as the work done by an atmospheric heat
engine with an efficiency of 1\%.  Preliminary simulations that we
have conducted indicate that a similar ratio is relevant for \hotjups\s
(Paper II).  If so, the implied kinetic energy generation 
is $10^2$ times
the intrinsic heat flux computed by current models.  Inclusion of this
energy could then lead to a first-order alteration in the behavior
predicted in evolution models.

In steady state, the kinetic energy that is produced
must be dissipated.  On Earth, this dissipation mostly
results from friction with the surface (Peixoto and Oort \cite{PO92},
p. 385).
For \hotjups, Kelvin-Helmholtz instabilities and breaking of gravity
and planetary waves are more relevant.  The key question is whether
the energy is dissipated in the ``weather'' layer, where starlight is
absorbed and radiation to space occurs, or in the deeper atmosphere.
In the former case, the dissipation will provide only an order-1\%
perturbation to the vertical profile of absorbed starlight and radiated thermal energy.
In the latter case, it comprises a hundred-fold alteration in the interior
energy budget.  We therefore need to know (i) to what pressures
can the energy be transported, and (ii) how deep must it be transported
to cause a major effect on the interior?

As discussed in Paper~II, mechanisms for transporting kinetic energy into the
interior include Kelvin-Helmholtz instabilities, direct vertical
advection, and waves.  The dynamical coupling between atmospheric layers suggests that
winds should develop throughout the radiative region even though the radiative cooling
and heating occurs predominantly at pressures less than a few bars.
The boundary between the radiative region and the convective
interior (at 100--1000 bars depending on the model) is a likely
location for dissipation, because Kelvin-Helmholtz instabilities and breaking
of downward propagating waves can both happen there.  Furthermore,
application of
the Taylor-Proudman theorem to the convective interior suggests that
winds should
develop throughout the convective interior even if the forcing occurs
only near the top
of the convective region.  This increases the possibility of dissipation in
the interior.

With the inclusion of an internal dissipative source, the energy
equation becomes
\begin{equation}
{\partial L \over \partial m} = \dot{\epsilon} - T{\partial S\over \partial
t},
\end{equation}
where $m$ is the mass inside any given level, and $\dot{\epsilon}(m)$ is the
energy dissipated per unit time per unit mass at that level.

The evolution of \hotjups\s including energy
dissipation has been studied by Bodenheimer et al. (\cite{Bod01}) in
the context of the tidal circularization of the orbit of the
planet. These authors focused on simulations where $\dot{\epsilon}$ was 
constant with $m$ (although they also performed some
simulations with spatially-varying dissipation).
The major difficulty is, as noted by the authors, the fact
that the present eccentricities of extrasolar planets within 0.1 AU of
their star are small and that the detected \hotjups\s generally do 
not possess close massive planetary companions which would impose on 
them a forced eccentricity.

Instead, we have argued that kinetic energy, generated from a portion of
the absorbed stellar flux, is transported to the interior
where it can be dissipated. Although the depth of such dissipation is unknown,
the majority could be deposited within
the radiative zone rather than throughout the interior.  Due to the
rapid rise of the Rosseland opacity with pressure and temperature, the effect
of heating anywhere within the convective core
is essentially equivalent to the case where it occurs entirely at the center
(a result shown by Bodenheimer et al.). The question is whether even shallower
heating --- say that occurring at tens to hundreds of bars, where atmospheric 
kinetic-energy deposition is likely --- can affect 
the evolution.  Therefore, we here explore
the influence of the dissipation's depth dependence and magnitude 
$\dot{E}=\int \dot{\epsilon} dm$.

An {\it ad hoc}, but reasonable, assumption is that a
fraction of up to 1\%
of the absorbed stellar flux is dissipated inside
the planet.  Quantitatively, we use $\dot{E}=2.4\times
10^{27}\rm\,erg\,s^{-1}$. Relatively small values of $\dot{E}$ can
affect the evolution, provided they are comparable or larger than the
luminosity obtained without dissipation $L$ (note that $L\sim
10^{24}-10^{25}\rm\,erg\,s^{-1}$) and affect the radiative gradient on
a sufficiently extended region of the interior. 

\begin{figure}[ht]
\begin{center}
\resizebox{\hsize}{!}{\includegraphics{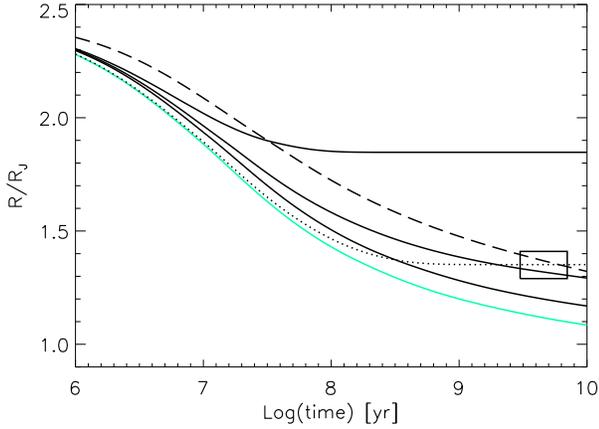}}
\caption{%
Evolution tracks obtained in the ``cold'' case, showing the influence
of dissipation. The bottom grey line corresponds to the case with
no dissipation. The other solid lines have been calculated
including the dissipation of 1\% of the absorbed stellar flux
($2.4\times
10^{27}\rm\,erg\,s^{-1}$), at various depths: from bottom to top,
dissipation was supposed to occur in various mass shells:
$\tilde{m}_0=10^{-5}$,
$2\times 10^{-5}$, or at the center of planet, respectively.
The dashed line corresponds to dissipation of 10\% of the absorbed
stellar flux ($2.4\times 10^{28}\rm\,erg\,s^{-1}$) with
$\tilde{m}_0=5\times 10^{-6}$. The dotted evolution track is for
dissipation of $\dot{E}=1.8\times 10^{26}\rm\,erg\,s^{-1}$ at the
planet's center. In the first two cases, energy dissipation occurs
mostly from the upper boundary to the nearly-isothermal region. 
The $\tilde{m}_0 = 5\times10^{-6}$, $10^{-5}$ and $2\times10^{-5}$
values of adimensional mass correspond to pressures of 5, 11,
and 21 bars, respectively.
}
\label{fig:evtracks}
\end{center}
\end{figure}

A first calculation assumes that energy is dissipated entirely at the
center of the planet. In that case, as shown in
Fig.~\ref{fig:evtracks} (uppermost solid line), an equilibrium with
the star is
reached after only $\sim 100$\,Ma, at which point the planet's radius
is 1.87\,\rjup\ and its structure remains unchanged with time (as long
as the star is also in equilibrium). This is very similar to the
results obtained by Bodenheimer et al. (2001). Also in agreement with
their results, we find that a calculation with the same
$\dot E$, but with the dissipation evenly distributed throughout the interior
(i.e., uniform $\dot \epsilon$) yields a curve similar to the upper
solid line in Fig.~\ref{fig:evtracks}.

In order to estimate the consequences of energy dissipation occurring closer
to the planet's surface, we use the following arbitrary
functional form:
\begin{equation}
\dot{\epsilon}=\dot{\epsilon}_0 e^{-(1-\tilde{m})/\tilde{m}_0},
\label{eq:epsilon}
\end{equation}
where $\tilde{m}$ is the adimensional mass (0 at planet's center and
1 at its surface), $\tilde{m}_0$ is the mass
fraction of the external regions over which most of the dissipation
occurs, and $\dot{\epsilon}_0$ is chosen such that
$\int\dot{\epsilon}dm=\dot{E}=2.4\times 10^{27}\rm\,erg\,s^{-1}$.
We will use values of $\tilde{m}_0$ equal to $5\times10^{-6}$, $10^{-5}$,
and $2\times10^{-5}$, which correspond to locations where the pressure
is 5, 11, and 21 bars, respectively.

A choice of $\tilde{m}_0=10^{-5}$ (which implies dissipative heating
distributed dominantly from the top boundary to 15\,bar but with a tail of
heating reaching $\sim 100$\,bar) yields a radius which
is, after a
few billion years, about 10\% larger than in the case with no
dissipation (third solid line from the top in Fig.~\ref{fig:evtracks}).
This is, in the ``cold'' case, insufficient to reproduce
the observed radius of HD209458b. A slightly higher value of
$\tilde{m}_0=2\times 10^{-5}$ yields an evolution track which is in
agreement with the
measured radius, as shown in Fig.~\ref{fig:evtracks} (second solid line from the
top). In that case,
the value of $1-\tilde{m}=\tilde{m}_0$ corresponds, for the model with
1.35\,\rjup\ to a pressure level $P=28\,$bar and $T=2800$\,K. However,
because of the form of Eq.~(\ref{eq:epsilon}), dissipation becomes
negligible only around $1-\tilde{m}\approx 10 \tilde{m}_0$,
i.e. $P=130$\,bar and $T=3380$\,K.

Two other evolution tracks have been calculated specifically to
illustrate how HD209458b's radius can be reproduced with different
values of $\dot E$ and $\tilde{m}_0$. In the case of dissipation at
the center, we were able to match an
equilibrium radius of 1.35\,\rjup\ with $\dot
E=1.8\times 10^{26}\rm\,erg\,s^{-1}$ (dotted line in
Fig.~\ref{fig:evtracks}), which is only 0.08\% of the global-mean absorbed
stellar flux. In the case of dissipation limited to a
shallow layer ($\tilde{m}_0=5\times 10^{-6}$, corresponding to a pressure
of 5 bars), we found that a
relatively high value of $\dot E$ corresponding to 10\% of the
absorbed stellar flux was necessary for the planet to contract to its
present radius in about $\sim 5$\,Ga (dashed line in
Fig.~\ref{fig:evtracks}).

\begin{figure}[ht]
\begin{center}
\resizebox{\hsize}{!}{\includegraphics{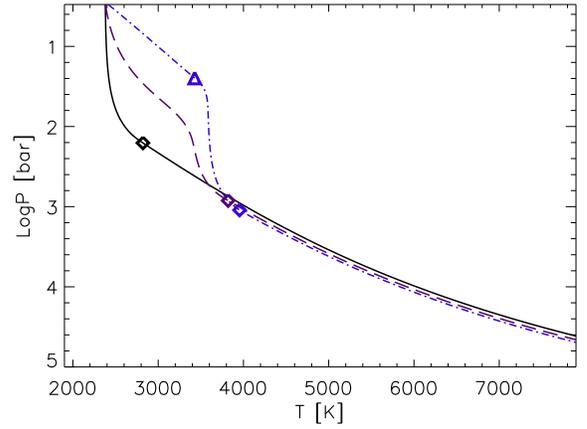}}
\caption{%
Temperature-pressure profiles for models of HD209458b (with
$R=1.35$\,\rjup) calculated with the ``cold''
atmospheric boundary condition, and assuming various dissipation
profiles: $\dot{E}=1.8\times
10^{26}\rm\,erg\,s^{-1}$ at the planet's center (solid line),
$\dot{E}=2.4\times 10^{27}\rm\,erg\,s^{-1}$ (1\% of the globally-averaged
absorbed stellar heat flux) and
$\tilde{m}_0=2\times 10^{-5}$ (dashed line), $\dot{E}=2.4\times
10^{28}\rm\,erg\,s^{-1}$ and
$\tilde{m}_0=5\times 10^{-6}$ (dot-dashed line). The diamonds
indicate the radiative/convective boundary. In the case of the
dot-dashed models, the large dissipation in the external layers is
responsible for a second convective zone pressures lower than 25 bars,
as indicated by a triangle.
}
\label{fig:profils}
\end{center}
\end{figure}

The temperature pressure profiles of three models calculated with the
``cold'' atmospheric boundary conditions but with different values of
the dissipation factor, and such that their total radius is
1.35\,\rjup\ , are compared in Fig.~\ref{fig:profils}. The temperature
profile of the model with dissipation at the center (solid line) is
essentially indistinguishable from our reference ``cold'' model which
included no dissipation but had a high intrinsic heat flux due to its
young age. As dissipation is increased but at the same time limited to
shallower outer layers, the temperature profile becomes more similar
to the ``hot'' case shown in Fig.~\ref{fig:prof_hotcold}. In the case of
the highest dissipation considered here but with a small
$\tilde{m}_0=5\times 10^{-6}$, an external (detached) convective region can
form. Note that in this case $\tint=790\,$K; atmospheric
models calculated with these high intrinsic effective temperatures
are also found to possess a deep convective zone (Barman et
al. \cite{Bar01}).

\subsection{Penetration of the stellar flux into the deep atmosphere}
\label{sec:flux}

We have shown that a kinetic energy flux corresponding to a small fraction of
the stellar flux can, if dissipated deep enough,
significantly affect the planet's evolution. The result would be
exactly the same if the stellar flux was radiatively transported to these
deep levels. We argue that stellar heat cannot be deposited so deep, however.

As shown by figures~\ref{fig:prof_hotcold} and \ref{fig:profils}, the
temperature profile
of HD209458b must cross the point defined by $P_{\rm i}\approx 1$\,kbar,
$T_{\rm i}\approx4000$\,K, assuming that the planet is of solar composition
(the addition of heavy elements tends to increase the temperature
required at a given pressure, but doesn't otherwise alter the
conclusions that follow). This point and the external boundary
condition then define the intrinsic luminosity required to reproduce
the measured radius, i.e.
\begin{equation}
\lint\approx {64\pi\sigma G M\over
3}{<T^3>\over<\kappa>}\left({dT\over dp}\right)_{\rm rad},
\end{equation}
where $\sigma$ is the Stefan-Boltzmann constant, $<T>$ and $<\kappa>$
are a mean temperature and opacity, respectively, and $(dT/dp)_{\rm rad}$
is the mean temperature lapse rate in the radiative zone.

In the ``hot'' case, the value of $\lint$ thus derived is small
because the difference in temperature between the bottom of the
radiative zone and the external boundary is small. This implies that
HD209458b needs to be relatively old to have such
a low intrinsic luminosity. In the more realistic ``cold'' case, the
planet either has to be uncomfortably young, or some additional heat
has to be transported to these levels. This requires
$\dot{\epsilon}> 0$, but with the additional requirement that the
temperature should be brought close to $\sim 3200$\,K at a pressure
$P<P_{\rm i}$.

By definition of the optical depth $\tau_\lambda$, the proportion of
stellar flux still remaining at a given level is equal to
$e^{-\tau_\lambda}$, where
\begin{equation}
\tau_\lambda=\int_z^\infty \kappa_\lambda \rho dz,
\label{eq:depth}
\end{equation}
and $\kappa_\lambda$ is the monochromatic opacity, and $z$ is
the altitude in the atmosphere. Equation~(\ref{eq:depth}) is strictly
valid only in a one-dimensional approximation, but this is
sufficient here. It is useful to approximate the integral by assuming
a constant opacity and density in a slice of atmosphere of height
equal to the pressure scale height. In that case, the pressure at
which 99\% of the stellar flux has been absorbed
can be estimated by:
\begin{equation}
P \wig{<} 5 {g\over \kappa_{\rm min}},
\end{equation}
where $\kappa_{\rm min}$ corresponds to the minimum value of the
opacity $\kappa_\lambda$. 

We estimate from the previous section that a penetration of 1\% of the
stellar flux to $P\sim 100$\,bar in the ``cold'' case allows the radius
of HD209458b to be explained without any other energy
dissipation. 99\% of the flux of a
6000\,K black body is emitted between 0.22 and 4.9\,$\mu$m. The
measured radius and mass implies that $g\approx 10^3\,\rm cm\,s^{-2}$,
therefore requiring $\kappa_{\rm min} \wig{<} 5\times 10^{-5}\rm\,
cm^2\,g^{-1}$.

This opacity is approximately the minimum expected for a pure
hydrogen-helium mixture at $P\sim 1$\,bar and $T\sim 1500$\,K, at
$\lambda\sim 1\,\mu$m due to Rayleigh scattering by H$_2$ and
H$_2$-H$_2$ collision-induced absorption (see e.g. Lenzuni et
al. \cite{Len91}; Guillot et al. \cite{Gui94}).
At temperatures above about 2000\,K, two very important sources of
continuous opacity arise, led by the increasing number of free
electrons: the free-free absoption of H$_2^-$ and the bound-free
absorption of the H$^-$ ion. However the number of free electrons in a
zero-metallicity gas remains low even at 3000\,K, and the low opacity
minimum persists to temperatures exceeding 3000\,K, and pressures exceeding
10\,bar. In this case a deep absorption of the stellar flux would then
be likely.

However, in a mixture of solar-like composition, a large fraction
of the electrons can be provided  by alkali metals.
Using electrons number densities obtained from Kurucz (\cite{Ku70}) and
Lodders (personal communication, 2001), 
we estimate the minimum continuous opacity
to climb to $3\times 10^{-3}\,\rm cm^2\,g^{-1}$ at 2500\,K and to $0.1\,\rm
cm^2\,g^{-1}$ at 3500\,K, mostly due to H$^-$ absorption (John
\cite{Jo88}). This alone
prevents any relevant fraction of the stellar flux to reach
levels at which the temperature is larger than 2500\,K. 

Furthermore, a number of other opacity sources are expected to
occur and even dominate the spectrum. Likely candidates are
K and Na which are now known to contribute
significantly to the atmospheric absorption of brown dwarfs with
similar temperatures, at visible wavelengths (Burrows et al.
\cite{BMS00}). Similarly, TiO is expected to provide an even larger
absorption at short wavelength where it appears in the deeper atmosphere.
For example, Barman et al. (\cite{Bar01}) find
that $\tau_{1.2\,\mu\rm m}=10$ is attained at pressures smaller than
$\sim 6$\,bar in the cloud-free atmosphere of \hotjups.
Finally, clouds, if present, would cause an absorption of
the stellar flux at even lower pressures.

It hence appears that only a zero-metallicity atmosphere would have a
low-enough opacity to allow the stellar flux to penetrate to
$P\sim 100$\,bar. This is an unlikely possibility, the metallicity of
HD209458 being close to solar (Mazeh et al. \cite{Maz00}).
One possibility remains however: that alkali metals and strong
absorbers such as TiO are buried deep due to condensation effects on the
night side (see Paper~II), so that the atmosphere on the day side
would be almost metal-free.
It is not clear even in this case that the measured radius could
be explained, because the lower overall opacities would increase
the rate of cooling and hence contraction of the planet.

In all the cases considered here, it seems very difficult for the
incoming stellar flux to penetrate down to levels where
the temperature is large (more than $\sim 2500$\,K). In order to
reproduce HD209458b's large radius, a temperature $\sim 4000\,K$ at a
pressure $P\sim 1\,$kbar must be attained.
Energy dissipation due to a transfer of kinetic energy
hence appears as the most likely missing energy source.

\section{Conclusions}

We have shown that the evolution of \hotjups\s is mainly driven by
processes occuring in their atmosphere and is consequently complex.
The measurement of the radius of one of these objects, HD209458b, has
allowed us to probe some of these mechanisms in detail.

We demonstrated that radiative-equilibrium atmospheric models predicting 
temperatures above
$\sim 2500$\,K at pressures $P\sim 10$\,bar are unlikely given the
rapid rise of the absorption with increasing temperature.
Cooler temperatures are to be expected in the atmosphere and
without other means than radiation to transport the incoming heat
flux, HD209458b's large radius cannot be reproduced unless the planet
is much younger than is revealed by observations of its parent star.

We showed the atmospheric temperature
variations to have a small effect on the planetary cooling, if limited
to a few 100's K.  The temperature variations lead to
faster cooling of the planet compared to standard models, which assume
the stellar heat to be evenly distributed onto the planet's
atmosphere. This accentuates the problem of reproducing HD209458b's radius.

Energy dissipation is however a very promising candidate to explain
HD209458b's missing heating source.  Lubow et al. (1997) have shown that tidal
synchronization of \hotjups\s could give rise to a large heat flux.
But this mechanism is limited to the early evolution of the planet and
should rapidly become negligible. Bodenheimer et
al. (\cite{Bod01}) argued that internal heating could be provided by tidal
circularization of an eccentric orbit. This is similarly
unlikely to occur in most \hotjups\s in the absence of a detected
close, massive companion capable of exciting their eccentricity. The
mechanism that we invoke is simply a downward transport of kinetic energy 
generated by the intense atmospheric heat engine.
We showed that only $\sim$0.08\% of the
stellar flux has to be transported to the interior regions to explain
the radius of HD209458b. This fraction rises to 1\% if heat
dissipation occurs predominantly in the outer $2\times 10^{-5}$ in
mass (reaching down to $\sim 2\times 10^{-4}$), or to
10\% if it occurs predominantly in the outer $5\times10^{-6}$
(reaching down to $\sim5\times 10^{-5}$).  Data for Earth show that 1\%
of the absorbed solar radiation is converted to kinetic energy and dissipated
in the atmosphere, and 1\% is plausible for \hotjups\s too.  To alter
the evolution, the energy need be deposited only a few scale heights below the
altitude where it is created, lending plausibility to the idea.

The presence of energy dissipation may be quantified in the future when
several \hotjups\s have been characterized. With several
ground programs (STARE, VULCAN), accepted space missions (COROT, MONS,
MOST) and proposed ones (KEPLER, EDDINGTON) aiming at detecting
photometric transits of \hotjups, there is indeed a good chance
that enough statistical information on the mass radius relationship of
\hotjups\s can be gathered. 

An unfortunate consequence of this study is that the possibility to
determine the planets' compositions solely from their
mass, radius and orbital characteristics seems to be postponed to a
more distant future. On the other
hand, we should rejoice over the perspective
of better understanding of irradiated atmospheres and tidal dissipation.
As usual, progress will mainly occur through observations and the
direct characterization of \hotjups.

\begin{acknowledgements}
This work benefited from many interactions with M.S. Marley, and
discussions over many years with members of the ``Tucson group''
(W.B. Hubbard, J.I. Lunine, A. Burrows). 
We also wish to thank P. Bodenheimer, R. Freedman,
K. Lodders, and D. Saumon for a variety of
useful contributions. 
This research was supported by the French {\it Programme National de
Plan\'etologie}, Institute of Theoretical Physics (NSF PH94-07194),
and National Research Council of the United States.
The numerical results described in this article are available from the
following URL: http://www.obs-nice.fr/guillot/pegasi-planets.
\end{acknowledgements}

\def\bi#1#2#3#4#5#6{{#1}, {#6}, {#3} {#4}, #5}


\begin{thebibliography}{}

\bibitem[1994]{AF94}
Alexander D.R., Ferguson J.W. 1994, \apj, 437, 879

\bibitem[2001]{Bar01}
Barman T., Hauschildt P.H., Allard F. 2001, \apj, 556, 885

\bibitem[2000]{Bod00}
Bodenheimer P., Hubickyj O., Lissauer J.J. 2000, Icarus, 143, 2

\bibitem[2001]{Bod01}
Bodenheimer P., Lin D.N.C., Mardling R. 2001, \apj, 548, 466

\bibitem[2001]{Betal01}
Brown T.M., Charbonneau D., Gilliland R.L., Noyes R.W.,
Burrows A. 2001, \apj, 552, 699

\bibitem[1997]{Bur97}
\bi{Burrows A., Marley M.S., Hubbard W.B., et al.}{A nongray theory
of extrasolar giant planets and brown dwarfs.}{\apj}{491}{856}{1997}

\bibitem[2000a]{Bur00}
Burrows A., Guillot T., Hubbard W.B., Marley M.S., Saumon D., Lunine J.I.,
Sudarsky D. 2000a, \apjl, 534, L97

\bibitem[2000b]{BMS00}
Burrows A., Marley M.S., Sharp C.M. 2000b, \apj, 531, 438

\bibitem[2000]{Cal00}
\bi{Charbonneau D., Brown T.M., Latham D.W., Mayor,
M.}{Detection of planetary transits across a Sun-like
star}{\apj}{529}{L45}{2000}

\bibitem[2000]{Gouk00}
\bi{Goukenleuque C., B\'ezard B., Joguet B., Lellouch E., Freedman
R.}{A radiative equilibrium model of 51 Peg b}{\icarus}{143}{308}{2000}

\bibitem[1999]{Gui99}
\bi{Guillot T.}{Review: Interiors of giant planets inside and
outside the solar system}{\science}{286}{72}{1999}

\bibitem[1994]{Gui94}
Guillot T., Gautier D., Chabrier G., Mosser B. 1994, {\it Icarus}
112, 337

\bibitem[1995]{Gui95}
Guillot T., Chabrier G., Gautier D., Morel P. 1995, \apj, 450, 463

\bibitem[1996]{Gui96}
\bi{Guillot T., Burrows A., Hubbard W.B., Lunine J.I.,
Saumon D.}{Giant planets at small orbital
distances}{\apj}{459}{L35}{1996}

\bibitem[1995]{GM95}
Guillot T., Morel P. 1995, A\&AS 109, 109

\bibitem[2000]{Hal00}
\bi{Henry G.W., Marcy G.W., Butler R.P., Vogt,
S.S.}{A transiting ``51 Peg-like'' planet}{\apj}{529}{L41-L44}{2000}

\bibitem[1977]{Hub77}
Hubbard W.B. 1977, Icarus, 30, 305

\bibitem[2001]{Hub01}
Hubbard W.B.,
Fortney J.J., Lunine J.I., Burrows A., Sudarsky D., Pinto P.A. 2001,
submitted to ApJ

\bibitem[1978]{IP78}
Ingersoll A.P., Porco C. 1978, Icarus

\bibitem[1988]{Jo88}
John T.L. 1988, \aa 193, 189

\bibitem[1970]{Ku70}
Kurucz R.L. Smithsonian Obs. Spec. Rep. 309, 1

\bibitem[1991]{Len91}
Lenzuni P., Chernoff D.F., Salpeter E.E. 1991, ApJS, 76, 759

\bibitem[1997]{Lub97}
Lubow S.H., Tout C.A., Livio M. 1997, \apj, 484, 866

\bibitem[1997]{Mar97}
Marcy G.W., Butler R.P., Williams E., Bildsten L., Graham J.R.,
Ghez A.M., Jernigan J.G., 1997, ApJ, 481, 926

\bibitem[2000]{Mar00}
Marcy G.W., Cochran W.D., Mayor M. 2000, Protostars and Planets IV
(Tucson: University of Arizona Press; eds Mannings V., Boss A.P.,
Russell S.S.), p. 1285

\bibitem[1996]{Mar96}
\bi{Marley M.S., Saumon D., Guillot T., et al.}{Atmospheric,
evolutionary and spectral models of the brown dwarf Gliese 229
B.}{\science}{272}{1919}{1996}

\bibitem[1995]{MQ95}
\bi{Mayor M., Queloz D.}{A Jupiter-mass companion to a
solar-type star}{\nature}{378}{355}{1995}

\bibitem[2001]{May01}
Mayor M., Naef D., Pepe F., Queloz D., Santos N., Udry S., Burnet M. 2001.
In Planetary Systems in the Universe: Observation, Formation and
Evolution, IAU Symp. 202, Eds. A. Penny, P. Artymowicz, A.-M. Lagrange
and S. Russel ASP Conf. Ser, in press

\bibitem[2000]{Maz00}
\bi{Mazeh T., Naef D., Torres G. et al.}{The spectroscopic orbit of
he planetary companion transiting HD 209458}{\apjl}{532}{L55}{2000}

\bibitem[1991]{Pea91}
\bi{Pearl J.C., Conrath B.J.}{The albedo, effective temperature, and
energy balance of Neptune, as determined from Voyager
data}{\jgr}{96}{18,921}{1991}

\bibitem[1992]{PO92}
\bi{Peixoto J.P., Oort A.H.} {} {\it Physics of Climate,}
{American Institute of Physics} {New York.} {1992}

\bibitem[1995]{Sau95}
Saumon D., Chabrier G. and Van Horn H.M. 1995. \apjs

\bibitem[1996]{Sau96}
Saumon D., Hubbard W.B., Burrows A., Guillot T., Lunine J.I., Chabrier G. 1996, \apj,
460, 993

\bibitem[2001]{SG01}
Showman A.P., Guillot T. 2001, submitted to A\&A (Paper II)

\bibitem[1998]{SS98}
\bi{Seager S., Sasselov D.D.}{Extrasolar giant planets
under strong stellar irradiation}{\apj}{502}{L157}{1998}

\bibitem[2000]{SS00}
Seager S., Sasselov D.D. 2000, \apj, 537, 916

\bibitem[1977]{SS77} 
Stevenson D.J., Salpeter E.E. 1977, \apjs, 35, 239 


\end{thebibliography}
\end{document}